\documentclass[pra,twocolumn]{revtex4}
\usepackage{graphicx}

\begin{document}
\title{Quantum entanglement and quantum nonlocality for $N$-photon entangled states}
\author{Yan-Hua Sun and Le-Man Kuang\footnote{Corresponding
author.}\footnote{Email address: lmkuang@hunnu.edu.cn (L. M.
Kuang)}}
\address{Department of Physics, Hunan Normal University, Changsha 410081, People's Republic of China}

\begin{abstract}
Quantum entanglement and quantum nonlocality of $N$-photon
entangled states $|\psi_{N m}\rangle
=\mathcal{N}_{m}[\cos\gamma|N-m\rangle_{1}|m\rangle_{2}
+e^{i\theta_{m}}\sin\gamma|m\rangle_{1}|N-m\rangle_{2}]$  and
their superpositions are studied. We indicate that the relative
phase $\theta_{m}$ affects quantum nonlocality but not quantum
entanglement for the state $|\psi_{N m}\rangle$. We show that
quantum nonlocality can be controlled and manipulated by adjusting
the state parameters of $|\psi_{N m}\rangle$, superposition
coefficients, and the azimuthal angles of the Bell operator. We
also show that the violation of the Bell inequality can reach its
maximal value  under certain conditions. It is found that quantum
superpositions based on $|\psi_{N m}\rangle$ can increase the
amount of entanglement, and give more ways to reach the maximal
violation of the Bell inequality.

\noindent PACS number(s): 03.65.Ud, 03.67.-a
\end{abstract}

\maketitle

\section{Introduction}
Quantum entanglement and quantum nonlocality are two striking
aspects of quantum mechanics. They are introduced by Einstein,
Podolsky, and Rosen (EPR) in their famous paper \cite{epr}. The
relationship between them has been paid much attention. They play
an essential role in the modern understanding of quantum
phenomena, and quantum information transmission and processing.
Bell proposed a remarkable inequality imposed by a local hidden
variable theory \cite{bel}, which enables a quantitative test on
quantum nonlocality. The quantum nonlocality test can be performed
on an entangled system composed of two coherent systems. This
entangled system can be used as a quantum entangled channel for
quantum information transfer. Numerous theoretical studies and
experimental demonstrations have been carried out to understand
nonlocal properties of quantum states. Various versions of Bell's
inequality \cite{chs,ch} are proposed. Gisin and Peres found pairs
of observable whose correlations violate Bell's inequality for a
discrete $N$-dimensional entangled state \cite{gis} . Banaszek and
W\'{o}dkiewicz studied Bell's inequality for continuous-variable
states in terms of Wigner representation in phase space based upon
parity measurement and displacement operation. Recently, Chen {\it
et al.} studied Bell's inequality of continuous-variable states
\cite{chen1} using their newly defined pseudospin Bell operators,
and they showed that the EPR state can maximally violate Bell's
inequality in their framework.

Recently, $N$-photon entangled states and their superposition
states are paid much attention.  They are widely used to realize
quantum lithography \cite{kuan,kok,bjo}, super-resolving phase
measurements \cite{mit}, and quantum teleportation \cite{coc}. In
this letter, we study quantum entanglement and quantum nonlocality
of the following $N$-photon entangled states and their
superpositions
\begin{eqnarray}
\label{e1}
|\psi_{N m}\rangle && =\mathcal{N}_{m}[\cos\gamma|N-m\rangle_{1}|m\rangle_{2} \nonumber \\
 &&+e^{i\theta_{m}}\sin\gamma|m\rangle_{1}|N-m\rangle_{2}],
\end{eqnarray}
where $m$ takes its values from zero to $N$, the normalization
factor is given by $\mathcal{N}_{m}^{-2}=1+\cos\theta_{m}\sin
2\gamma \delta_{N,2m}$ with $\gamma$ and  $\theta_{m}$ being an
entanglement angle and a relative phase, respectively. We will
calculate the von Neumann entropy and study the Bell's inequality
for quantum states $|\psi_{N m}\rangle$ and their superposition
states. We will show that quantum superpositions for the two modes
may increase the amount of entanglement, and $N$-photon entangled
states  can maximally violate Bell's inequality. This letter is
organized as follows. In Sec. II, we study quantum entanglement of
$N$-photon entangled states and their superposition states through
analyzing their von Neumann entropy.  Quantum nonlocality of
$N$-photon entangled states and their superposition states is
investigated through discussing the violation of the Bell's
inequality in the pseudospin Bell-operator formalism developed by
Chen {\it et al.} \cite{chen1} in Sec. III. The last section is
devoted to summary and conclusion.

\section{Quantum entanglement for $N$-photon entangled states}

In this section, we study properties of quantum entanglement of
the $N$-photon entangled states given by Eq. (\ref{e1}). In order
to this, we consider the nontrivial case of $N\neq 2m$, in which
the normalization constant $\mathcal{N}_{m}=1$. The degree of the
entanglement can be described by the von Neumann entropy defined
by
\begin{eqnarray}
\label{e2}
 E(\rho_{1})=-Tr_{1}(\rho_{1}log\rho_{1}),
\end{eqnarray}
where $\rho_{1}$ is the reduced density operator of the first
mode.

For the $N$-photon entangled states defined by Eq. (\ref{e1}) we
find the von Neumann entropy to be
\begin{eqnarray}
\label{e3}
 E_{1}&=&-\cos^{2}\gamma\log\cos^2\gamma -
\sin^{2}\gamma\log\sin^2\gamma,
\end{eqnarray}
which indicates that quantum entanglement of the $N$-photon
entangled states given by Eq. (\ref{e1}) is independent of the
superposition phases $\theta_m$ and the total photon number of the
two modes $N$, and changes periodically with respect to the
entanglement angle $\gamma$. In particular, the amount of
entanglement reaches the maximal value of $E_{1}=1$ when the
entanglement angle takes values by  $\gamma=k\pi + \pi/4$ with $k$
being an integer. In Fig. 1 we plot the change of the von Neumann
entropy with respect to the entanglement angle.

\begin{figure}[htp]
\center
\includegraphics[width=3.3in,height=2.1in]{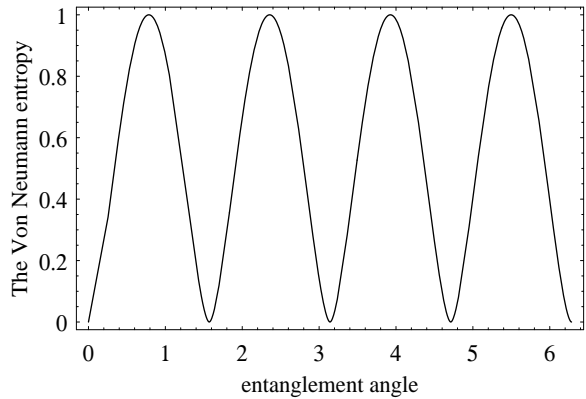}
\caption{ The von Neumann entropy of the $N$-photon entangled
state is plotted against the entanglement angle $\gamma$.}
\end{figure}

We then study quantum entanglement of superposition states based
on  the $N$-photon entangled states given by Eq. (\ref{e1}).
Firstly, we consider a two-state superposition state defined by
\begin{eqnarray}
\label{e4}
 |\Psi_{2}\rangle=\frac{1}{\sqrt{2}}(|\psi_{30}\rangle +
|\psi_{31}\rangle),
\end{eqnarray}
which leads to the von Neumann entropy
\begin{eqnarray}
\label{e5}
E_{2}=1-\cos^{2}\gamma\log\cos^2\gamma -
\sin^{2}\gamma\log\sin^2\gamma,
\end{eqnarray}
which implies that the amount of entanglement of the two-component
superposition state only depends  on the entanglement angle
$\gamma$ of its basis. Especially, comparing Eq. (\ref{e5}) with
(\ref{e5}) we find that the difference of the amount of
entanglement between the two-component superposition state
(\ref{e4}) and the basis state (\ref{e1}) is a positive constant,
i.e., $E_2-E_1=1$. This indicates that the entanglement amount of
the superposition state is always larger than that of the basis
state for an arbitrary entanglement angle $\gamma$. In other
words, starting with basis states defined by (\ref{e1}) one can
construct quantum superposition states with larger amount of
entanglement than that of the basis states. Hence, we may conclude
that quantum superpositions for the two modes may increase the
amount of entanglement.

In order to further demonstrate the above idea of quantum
superpositions increasing quantum entanglement, in what follows we
take into account a multi-component quantum superposition state
consisting of $N$ bases with a fixed photon number $N$, but with
different distributions $m$,
\begin{eqnarray}
\label{e6}
|\psi_{N}\rangle&=&A\sum_{m=0}^{N}\alpha_{m}|\psi_{Nm}\rangle,
\end{eqnarray}
where $A$ is a normalization factor. It is straightforward to
express the $N$-component superposition state as the following
number-sum Bell state,
\begin{eqnarray}
\label{e7}
|\Psi_{N}\rangle&=&\sum_{m=0}^{N}d_{m}|N-m\rangle_{1}|m\rangle_{2},
\end{eqnarray}
which is the eigenstate of the number-sum Bell operators
$\hat{N}=\hat{N}_1+\hat{N}_2$ with $\hat{N}_i$ being the number
operators of the respective modes.  The coefficients in Eq.
(\ref{e7}) are given by
\begin{eqnarray}
\label{e8}
d_{m}=A(\alpha_{m}\mathcal{N}_{m}\cos\gamma+
\alpha_{N-m}\mathcal{N}_{N-m}e^{i\theta_{N-m}}\sin\gamma),
\end{eqnarray}
where the normalization of the superposition state (\ref{e8})
implies that $d_{m}$ satisfy the condition
$\sum_{m=0}^{N}|d_{m}|^{2}=1$.

In order to calculate the von Neumann entropy of the superposition
state (\ref{e7}), we need the reduced density operator
\begin{eqnarray}
\label{e9}
\rho_{1}&=&\sum_{m=0}^{N}|d_{m}|^{2}|N-m\rangle_{1}\langle
N-m|,
\end{eqnarray}
which leads to the following von Neumann entropy
\begin{eqnarray}
\label{e10} E_{N}&=&-\sum_{m=0}^{N}|d_{m}|^{2}\log|d_{m}|^{2}.
\end{eqnarray}

\begin{figure}[htp]
\center
\includegraphics[width=3.5in,height=2.1in]{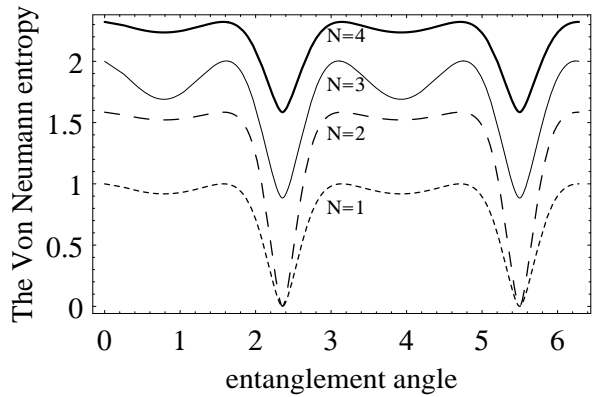}
\caption{ The von Neumann entropy of  multi-component
superposition states based on $|\psi_{Nm}\rangle$ is plotted
against the entanglement angle $\gamma$ for $N=1, 2, 3$, and $4$,
respectively.}
\end{figure}

For the sake of simplicity and without the loss of generality, we
consider the situation of equal-weight superposition in which we
choose the superposition coefficients $\alpha_{m}=1/\sqrt{N+1}$
and the relative phases $\theta_{m}=2\pi m/N$. In this case, we
have
\begin{eqnarray}
\label{e11}
|d_{m}|^{2}=A^2|\mathcal{N}_{m}|^{2}\{1+\cos[2\pi(N-m)/N]\sin2\gamma\},
\end{eqnarray}
where the normalization constant is given by
\begin{eqnarray}
\label{e12}
A^{-2}=\sum_{m=0}^{N}|\mathcal{N}_{m}|^{2}\{1+\cos[2\pi(N-m)/N]\sin2\gamma\}.
\end{eqnarray}

Then the von Neumann entropy of the superposition state can be
directly obtained through substituting (\ref{e11}) and (\ref{e12})
into (\ref{e10}). From Eqs. (\ref{e10})-(\ref{e12}) we can find
the maximal value of the von Neumann entropy of the superposition
state to be $E_{N,max}=\log_2(N+1)$ when $\gamma=k\pi/2$ with $k$
being an arbitrary integer. This implies that the maximal amount
of entanglement of the $N$-component  superposition state only
depends on the total photon number $N$, and increases with
increasing the total photon number $N$. In order to clearly see
the influence of the number of components $N$ and the entanglement
angle $\gamma $, in Fig. 2 we plot the von Neumann entropy of the
superposition states when $N=1, 2, 3$, and $4$, respectively. From
Fig. 2 we can see that the entanglement amount increases with
increasing  the number of components, i.e., the photon number $N$,
and changes  periodically with respect to the entanglement angle
$\gamma $.

\section{Quantum nonlocality  for $N$-photon entangled states}

 In this section, we study quantum nonlocality of $N$-photon entangled states and
their superposition states through discussing the violation of the
Bell's inequality in the pseudospin Bell-operator formalism
developed by Chen and coworkers \cite{chen1}. Let us begin with a
brief review of the pseudospin-operator formalism \cite{chen1}.
For a single-mode boson field, the pseudospin operators can be
defined in terms of project operators in a Fock space in the
following form
\begin{eqnarray}
\label{e13} S_{z}&=&\sum_{n=0}^{\infty}[| 2n+1 \rangle\langle 2n+1
|-|2n\rangle\langle 2n
|],\nonumber\\
S_{-}&=&\sum_{n=0}^{\infty}[| 2n \rangle\langle 2n+1
|]=(S_{j+})^{\dag},
\end{eqnarray}
where  $ |n\rangle$ are the usual Fock states of the boson mode.
The operator $S_{z}=-(-1)^{\hat{N}}$ with $\hat{N}$ being the
number operator and $(-1)^{N} $ being the parity operator, $S_{+}$
and $S_{-}$ being the ``parity-flip" operators. They satisfy  the
commutation relations of the $su(2)$ Lie algebra
\begin{eqnarray}
\label{e14}
 [S_{z},S_{ \pm}]=\pm 2S_{\pm},\hspace{0.3cm}
[S_{+},S_{-}]&=S_{z}.
\end{eqnarray}

For an  arbitrary  vector living on the surface of  a  unit sphere
$\vec{a}$ = ($\sin\theta_{a}\cos\varphi_{a},
\sin\theta_{a}\sin\varphi_{a}, \cos\theta_{a}$), we have the
following dot product
\begin{eqnarray}
\label{e15}
\vec{a}\cdot\vec{S}=S_{z}\cos\theta_{a}+\sin\theta_{a}(e^{i\varphi_{a}}S_{-}+e^{-i\varphi_{a}}S_{+}).
\end{eqnarray}

Then  for a two-mode boson field, the Bell operator due to
Clauser, Horne, Shimony, and Holt (CHSH) \cite{cla} can be defined
by
\begin{eqnarray}
\label{e16} B&=& (\vec{a}\cdot\vec{S_{1}}) \otimes
(\vec{b}\cdot\vec{S_{2}})+(\vec{a}\cdot\vec{S_{1}})\otimes
(\vec{b'}\cdot\vec{S_{2}}) \nonumber \\
&&+(\vec{a'}\cdot\vec{S_{1}})\otimes
(\vec{b}\cdot\vec{S_{2}})-(\vec{a'}\cdot\vec{S_{1}})\otimes
(\vec{b'}\cdot\vec{S_{2}}),
\end{eqnarray}
where $ \vec{a'}, \vec{b}$, and $\vec{b'}$ are three unit vectors
similarly defined as $\vec{a}$,  $\vec{S_{1}}$ and $\vec{S_{2}}$
are defined as in Eq. (\ref{e13}).

As well known, local hidden variable theories impose the Bell-CHSH
inequality $|\langle B \rangle|\leq 2$  where $\langle B \rangle$
is the mean value of the Bell operator with respect to a given
quantum state. However, in the quantum theory it is found that
$|\langle B \rangle|\leq 2\sqrt{2}$, which implies that the
Bell-CHSH inequality is violated. In particular, when $|\langle B
\rangle|= 2\sqrt{2}$ for a given quantum state, we say that the
Bell-CHSH inequality is maximally violated by the quantum state.

Quantum nonlocality of a quantum state can be described by the
violation of the Bell-CHSH inequality. The expectation value of
the Bell operator with respect to  a quantum state $|\psi \rangle$
can be expressed in terms of the correlation functions as
\begin{eqnarray}
\label{e17}\langle B
\rangle&=&E(\theta_{a},\theta_{b})+E(\theta_{a},\theta_{b'})+E(\theta_{a'},\theta_{b})\nonumber
\\&&-E(\theta_{a'},\theta_{b'}),
\end{eqnarray}
where the correlation functions are defined by
\begin{equation}
\label{e18} E(\theta_{a},\theta_{b})=\langle
\psi|S_{\theta_{a}}^{(1)}\otimes S_{\theta_{b}}^{(2)}|\psi\rangle,
\end{equation}
with
\begin{equation}
\label{e19}
S_{\theta_{a}}^{(j)}=S_{jz}\cos\theta_{a}+S_{jx}\sin\theta_{a}.
\end{equation}

We now investigate quantum nonlocality of  the $N$-photon
entangled state given by Eq. (\ref{e1}). For this multi-photon
entangled state we find the correlation function to be
\begin{eqnarray}
\label{e20}
E_{Nm}(\theta_{a},\theta_{b})&=&\mathcal{N}_{m}^{2}\{[(-1)^{N}+K(\theta_{m},\gamma)\delta_{N,2m}]\nonumber\\
&&\times\cos\theta_{a}\cos\theta_{b}
+\delta_{N,2m\pm 1}\nonumber\\
&&\times K(\theta_{m},\gamma)\sin\theta_{a}\sin\theta_{b}\},
\end{eqnarray}
where we have introduced the following effective state parameter
\begin{eqnarray}
\label{e21}K(\theta_{m},\gamma)=\cos\theta_{m}\sin 2\gamma,
\end{eqnarray}
which describes the effect of the basis state defined by
(\ref{e1}) on the correlation functions.

Making use of the correlation function (\ref{e20}), from Eq.
(\ref{e17}) we can get the expectation value of the Bell operator
for arbitrary values of all azimuthal angles $\theta_{a}$,
$\theta_{b}$, $\theta_{a'}$ and $\theta_{b'}$. As a concrete
example, we consider the situation of $\theta_{a}=0$,
$\theta_{a'}=\pi/2$ and $\theta_{b}=-\theta_{b'}$. In this case,
from Eqs. (\ref{e17}), (\ref{e20}) and (\ref{e21}) we can obtain
the expectation value of the Bell operator given by
\begin{eqnarray}
\label{e22}\langle B
\rangle&=&2\mathcal{N}_{m}^{2}\{[(-1)^{N}+K(\theta_{m},\gamma)\delta_{N,2m}]\cos\theta_{b}\nonumber\\
&&+\delta_{N,2m\pm 1}K(\theta_{m},\gamma)\sin\theta_{b}\}.
\end{eqnarray}

In order to observe quantum nonlocality of the $N$-photon
entangled state (\ref{e1}), we consider two different cases of
$N=2m$ and $N=2m \pm 1$, respectively. When $N=2m$, the quantum
state given by Eq. (\ref{e1}) is disentangled, and reduces to
\begin{eqnarray}
\label{e23}|\psi_{N
m}\rangle=\mathcal{N}_{m}(\cos\gamma+e^{i\theta_{m}}\sin\gamma)|m\rangle_{1}|m\rangle_{2}.
\end{eqnarray}

Making use of  Eqs. (\ref{e17}), (\ref{e18}) and (\ref{e20}), we
obtain the expectation value of the Bell operator with respect to
the disentangled state (\ref{e23}) $\langle B
\rangle=2\cos\theta_{b}$, which means that $|\langle B
\rangle|\leq 2$. Hence, the unentangled state (\ref{e23}) cannot
produce a violation of Bell's inequality.

On the other hand, when $N=2m\pm 1$, the expectation value of the
Bell operator  with respect to the state (\ref{e1}) is given by
\begin{eqnarray}
\label{e24}\langle B
\rangle=2[K(\theta_{m},\gamma)\sin\theta_{b}-\cos\theta_{b}],
 \end{eqnarray}
which indicates that for a fixed entanglement angle $\gamma$ and a
fixed relative phase $\theta_m$,  when $\theta_{b}=-\tan^{-1}K$,
the expectation value of the Bell operator $\langle B \rangle$
reaches its maximum given by
\begin{eqnarray}
\label{e25} \langle B
\rangle_{max}=2\sqrt{1+K^{2}(\theta_{m},\gamma)},
 \end{eqnarray}
which implies that $\langle B \rangle_{max}\geq 2$ due to
$|K(\theta_{m},\gamma)|\leq1.$ Thus, the $N$-photon entangled
state always violates the Bell's  inequality if
$K(\theta_{m},\gamma)\neq 0$.  From Eqs. (\ref{e21})
and(\ref{e24}) we can see that the degree of violation of the
Bell's inequalities depends upon  both the entangling angle
$\gamma$ and the relative phase $\theta_{m}$, and it changes
periodically with respect to both $\gamma$ and $\theta_{m}$. In
particular, we note that the relative phase $\theta_{m}$ seriously
affects the mean value of the Bell operator although it dos not
affect quantum entanglement of the quantum state given by
(\ref{e1}). It is easy to find that when $|\psi_{N m}\rangle$
reaches maximal entanglement i.e., $\gamma=\pi/4$, and
$\theta_{m}=0$,  we have $K(\theta_{m},\gamma)=1$. Under these
conditions,  we can reach the maximal violation of Bell's
inequality with $ \langle B \rangle_{max}=2\sqrt{2}$.

In Fig. 3 we plot the degree of the violation of the Bell's
inequality for $N$-photon entangled state against $\gamma$ and
$\theta_{m}$. From Fig. 3 we can see that the mean value of the
Bell operator with respect to the multi-photon entangled state (1)
changes periodically with both of the entanglement angle and the
relative phase, and the Bell's inequality is always violated for
arbitrary values of the  entanglement angle and the relative phase
except $\gamma=k\pi/2$ and $\theta=(2k+1)\pi/2$ with $k$ is an
integer.

\begin{figure}[htp]
\center
\includegraphics[width=3.3in,height=3.1in]{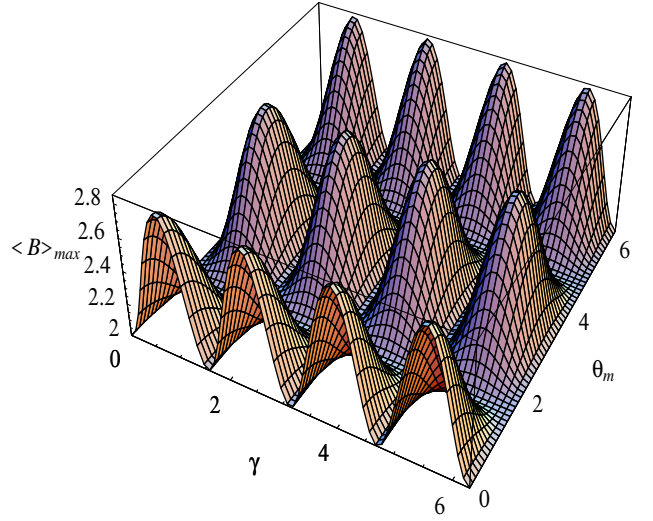}
\caption{The degree of the maximal violation of the Bell's
inequality for $N$-photon entangled state  $\langle B
\rangle_{max}$ is plotted against the entanglement angle $\gamma$
and the relative phase $\theta_{m}$. }
\end{figure}

Finally,  we  consider quantum nonlocality of quantum
superposition states which is formed using the multi-photon
entangled state (\ref{e1}). In order to calculate the expectation
value of the Bell operator with respect to the superposition
states based on entangled states given by (\ref{e1}), we need the
following correlation function
\begin{eqnarray}
\label{e26} E_{m m'}=\langle\psi_{N m}|S_{\theta_{a}}^{(1)}\otimes
S_{\theta_{b}}^{(2)}|\psi_{N m'}\rangle,
\end{eqnarray}
which is given by the following expression
\begin{eqnarray}
\label{e27}E_{m
m'}&=&\mathcal{N}_{m}\mathcal{N}_{m'}^{*}\cos\theta_{a}\cos\theta_{b}\nonumber
\\
&&\times \left \{(-1)^{N}\delta_{m,m'}\delta_{N,m+m'}\cos\gamma
\sin\gamma  \right.\nonumber\\
& &\times [\cos^{2}\gamma+e^{i (\theta_{m'}-\theta_{m})} \sin^{2}\gamma] \nonumber\\
&& + (-1)^{m+m'}(e^{i \theta_{m'}} +e^{-i\theta_{m}})\}\nonumber\\
&& + \mathcal{N}_{m}\mathcal{N}^*_{m'}\sin\theta_{a}\sin\theta_{b}
\nonumber \\
&& \times \left\{[\cos^{2}\gamma +
e^{i(\theta_{m'}-\theta_{m})}\sin^{2}\gamma]\delta_{m,m'\pm 1}
\right. \nonumber\\
&& \left. + \cos \gamma \sin
\gamma(e^{i\theta_{m'}}+e^{-i\theta_{m}})\delta_{N,m+m'\pm
1}\right \}.
\end{eqnarray}

As a simple example of analyzing quantum nonlocality of
superposition states, we consider the following two-component
superposition state,
\begin{eqnarray}
\label{e28}|\psi_{3}\rangle=C(\alpha_{0}|\psi_{30}\rangle+\alpha_{1}|\psi_{31}\rangle),
\end{eqnarray}
where $C$ is the normalization constant. Making use of Eq.
(\ref{e27}), we can obtain the expectation value of the Bell
operator for this superposition state
\begin{eqnarray}
\label{e29}\langle B \rangle&=&-2\cos\theta_{b} +
4\sin\theta_{b}(|\alpha_{0}|^{2}+|\alpha_{1}|^{2})^{-1}\nonumber
\\&&\times Re[\alpha_{0}^{*}\alpha_{1}(\cos^{2}\gamma +e^{i
(\theta_{1}-\theta_{0})} \sin^{2}\gamma)],
\end{eqnarray}
which indicates that $\langle B \rangle$ depends upon not only the
superposition coefficients $\alpha_{0},\alpha_{1}$, the azimuthal
angles $\theta_{a}, \theta_{b}$, but also upon the state
parameters of the basis states given by Eq. (\ref{e1}),
$\theta_{0}, \theta_{1}$ and $\gamma$.

In order to observe the maximal violation of the Bell inequality,
we can choose  $\alpha_{0}=\alpha_{1}=1$, the Bell function
$\langle B \rangle$  given by Eq. (\ref{e29}) becomes
\begin{eqnarray}
\label{e30}\langle B \rangle&=&2\sin\theta_{b}[\cos^{2}\gamma
+\cos(\theta_{1}-\theta_{0})\sin^{2}\gamma]\nonumber\\
&&-2\cos\theta_{b}.
\end{eqnarray}

Making use of Eq. (\ref{e30}), we can show  that the maximal
violation of the Bell inequality can be reached by controlling the
azimuthal angle $\theta_{b}$, the state parameters of the basis
states given by Eq. (1), $\theta_{0}, \theta_{1}$ and $\gamma$. In
fact, when $\theta_{1}-\theta_{0}=2k\pi$ with $k$ being an
arbitrary integer, we can arrive at the following Bell function
\begin{eqnarray}
\label{e31}\langle B \rangle&=2(\sin\theta_{b}-\cos\theta_{b})
\end{eqnarray}
which means that  the values of the Bell function of $\langle B
\rangle$ are independent of the entangling parameter of the basis
state defined in (1), and we can reach the maximal violation of
the Bell inequality $\langle B \rangle_{max}=2\sqrt{2}$  when
$\theta_{b}=-\pi/4$.

On the other hand, when the phase difference takes
$\theta_{1}-\theta_{0}=(2k+1)\pi$ with $k$ being an arbitrary
integer,  the Bell function given by Eq. (30) becomes
\begin{eqnarray}
\label{e32}\langle B
\rangle&=2[\sin\theta_{b}\cos(2\gamma)-\cos\theta_{b}],
\end{eqnarray}
from which we can see that when the entanglement angle satisfies
$\gamma=k\pi$ with $k$ being an arbitrary integer, Eq. (\ref{e32})
becomes Eq. (\ref{e31}), hence the maximal violation of the Bell
inequality is reached with $\theta_{b}=-\pi/4$. However, when the
entanglement angle satisfies $\gamma=(2k+1)\pi/2$ with $k$ being
an arbitrary integer, the Bell function (\ref{e32}) reduces to
\begin{eqnarray}
\label{e33}\langle B \rangle&=-2(\sin\theta_{b}+\cos\theta_{b}),
\end{eqnarray}
which implies that the violation of the Bell inequality reaches
its maximal value $\langle B \rangle_{max}=2\sqrt{2}$  when
$\theta_{b}=\pi/4$. Therefore, we can conclude that superposition
states based on entangled states (\ref{e1}) can provide us with
more ways of reaching the maximal violation of the Bell's
inequality.

\section{Concluding remarks}

In summary, we have studied quantum entanglement and quantum
nonlocality of  $N$-photon entangled states for the two modes
defined in Eq. (1) and their superpositions through investigating
the von Neumann entropy and the violation of the Bell's inequality
in the pseudospin Bell-operator formalism. For the multi-photon
entangled states defined in Eq. (1) we have indicated that the von
Neumann entropy is independent of the relative phase $\theta_{m}$.
Hence, quantum entanglement of these states only depends on the
entanglement angle. However, quantum nonlocality of these quantum
states exhibits different dependence upon the state parameters. We
have shown that both of the entanglement angle and the relative
phase seriously affect the violation of the Bell's inequality. And
we have indicated that under certain conditions the maximal
violation of the Bell's inequality can be reached.

It is worthwhile to mention that multi-component superposition
states made from $N$-photon entangled states for the two modes
defined in Eq. (1) exhibit some interesting characteristics on
their quantum entanglement and quantum nonlocality.  Firstly, we
have found that these multi-component superposition states have
larger amount of entanglement than that of the basis states
defined by (1). Hence, quantum superpositions for the two modes
can increase the amount of entanglement. This indicates the
possibility of obtaining entangled states with a larger amount of
entanglement starting from entangled states with a smaller amount
of entanglement. Secondly, we have found that for these quantum
superposition states there are more ways to make corresponding
Bell's inequality reach the maximal violation, and revealed that
quantum nonlocality can be controlled and manipulated by adjusting
the state parameters of $|\psi_{N m}\rangle$, superposition
coefficients, and the azimuthal angles of the Bell operator. We
hope that these results obtained in present paper would find their
applications in quantum information processing
\cite{nie,che,lu,zhou2} and the test of quantum nonlocality
\cite{jeo,pit}.

\acknowledgments This work was supported in part the National
Fundamental Research Program (2001CB309310), the National Natural
Science Foundation of China under Grant Nos. 90203018,
10325523,10347128 and 10075018, the foundation of the Education
Ministry of China, and the Educational Committee of Human Province
under Grant Nos. 200248 and 02A026.


\begin{references}
\bibitem{epr} A. Einstein, B. Podolsky, and N. Rosen, Phys. Rev. 47 (1935) 777.
\bibitem{bel} S. Bell, Physics (Long Island City, N.Y.) 1 (1964)195.
\bibitem{chs} J.F. Clauser, M.A. Horne, A. Shimony, and R.A. Holt, Phys. Rev. Lett. 23 (1969) 880.
\bibitem{ch}  J.F. Clauser and M.A. Horne, Phys. Rev. D 10 (1974) 526.
\bibitem{gis} N. Gisin and A. Peres, Phys. Lett. A 162  (1992) 15.
\bibitem{chen1} Z. Chen, J. Pan, G. Hou, and Y. Zhang, Phys. Rev. Lett. 88 (2002) 040406.
\bibitem{kuan} Y.H. Wang and L.M. Kuang, Journal of Optics B: Quantum and Semiclassical Optics 5 (2003) 405.
\bibitem{kok} P. Kok, A.N. Boto, D.S. Abrams, C.P. Williams, S.L. Braunstein, and J.P. Dowling, Phys. Rev. A 63 (2001)  063407.
\bibitem{bjo} G. Bj\"{o}rk, L. L. S\'{a}nchez-Soto, J. S\"{o}derholm, Phys. Rev. Lett. 86 (2001)  4516 .
\bibitem{mit} M.W. Mitchell, J.S. Lundeen and A. M. Steinberg, Nature  429 (2004) 161.
\bibitem{coc} P.T. Cochrane, G. J. Milburn, and W. J. Munro, Phys. Rev. A62 (2000)  062307.
\bibitem{nie} M. Nielsen  and I. L. Chuang, Quantum Computation and Quantum
Information (Cambridge University, Cambridge, England,2000 ).
\bibitem{che} Z. Chen, G. Hou, and Y. Zhang,  Phys. Rev. A  65 (2002)  032317.
\bibitem{lu} J. Lu, L. Zhou, and L. M. Kuang, Phys. Lett. A 330 (2004) 48.
\bibitem{zhou2} L. Zhou, L.M. Kuang, Phys. Lett. A 315 (2003) 426.
\bibitem{che} Z. Chen, G. Hou, and Y. Zhang,  Phys. Rev. A  65 (2002)  032317.
\bibitem{jeo} H. Jeong, W. Son, M.S. Kim, D. Ahm, and $\check{C}$. Brukner, Phys. Rev. A 67 (2003) 012106.
\bibitem{pit}I. Pitowsky and K. Svozil,  Phys. Rev. A  64 (2001)  014102.
\end{references}
\end{document}